\documentclass[12pt,a4paper]{article}
\usepackage[T2A]{fontenc}
\usepackage[utf8]{inputenc}
\usepackage[english]{babel}
\usepackage{amssymb,graphicx}
\usepackage{amsmath,amsfonts}
\usepackage{amsthm}
\usepackage{framed}
\usepackage{fullpage}
\usepackage{feynmf}
\usepackage{microtype}
\usepackage{hyperref}

 \newcommand{\tr}{\mathop{\mathrm{tr}}\nolimits}

\title{Quantum spectral curve for $(q,t)$-matrix model} \author{Yegor
  Zenkevich$^{a,b,c,d}$\thanks{yegor.zenkevich@gmail.com}\\
  {\small $^a$\textit{ITEP, Moscow, Russia}}\\
  {\small $^b$\textit{Dipartimento di Fisica, Universita di
      Milano-Bicocca, Milan, Italy}}\\
  {\small $^c$\textit{INFN, sezione di Milano-Bicocca, Milan, Italy}}\\
  {\small $^d$\textit{NRNU Moscow Engineering Physics Institute,
      Moscow, Russia}}} \date{}

\begin{document}
\maketitle

\vspace{-50ex}
\begin{flushright}
  ITEP-TH-12/15\\
  INR-TH/2015-015
\end{flushright}
\vspace{38ex}

\begin{abstract}
  We derive quantum spectral curve equation for $(q,t)$-matrix model,
  which turns out to be a certain difference equation. We show that in
  Nekrasov--Shatashvili limit this equation reproduces the Baxter TQ
  equation for the quantum XXZ spin chain. This chain is spectral dual
  to the Seiberg--Witten integrable system associated with the AGT
  dual gauge theory.

  MSC classes: 39A13, 15B52, 81T75, 81T40

  Keywords: matrix models, quantum spectral curves, conformal field
  theory, integrable models
\end{abstract}

\section{Introduction}
\label{sec:introduction}

Quantum generalizations of classical spectral curves have attracted
much attention recently. They appear in a variety of different
contexts in mathematical physics, some of them interrelated. In
topological strings quantum spectral curves are the equations solved
by the topological recursion
procedure~\cite{top-rec1},~\cite{top-rec2},~\cite{top-rec3},~\cite{top-rec4},~\cite{top-rec5},~\cite{top-rec6},~\cite{top-rec7},~\cite{top-rec8},~\cite{top-rec9}. In
$2d$ CFT they provide the differential equation for conformal block
with a degenerate field insertion~\cite{degen-field}. In
$\mathcal{N}=2$ gauge theories quantum spectral curves describe
Nekrasov partition functions with extra surface
operators~\cite{math1},~\cite{math2},~\cite{surf-opers1},~\cite{surf-opers2}. In
quantum integrable systems they appear as Baxter TQ equations or
universal difference operators~\cite{baxter-eqs}. Finally in knot
theory similar equations provide recurrence relations for colored
HOMFLY polynomials in different
representations~\cite{a-poly1},~\cite{a-poly2},~\cite{a-poly3},~\cite{a-poly4}.

In this short letter we derive the quantum spectral curve for the
$(q,t)$-matrix model. It is related to all of the fields mentioned
above. Firstly, the $(q,t)$-matrix models (or the $q$-deformed
$\beta$-ensemble) provide the Dotsenko--Fateev representation for the
$q$-deformed Virasoro conformal
blocks~\cite{q-DF1},~\cite{q-DF2},~\cite{q-DF3}. Due to the AGT
correspondence~\cite{AGT1},~\cite{AGT2},~\cite{AGT3},~\cite{AGT4},
these blocks are related to five dimensional gauge
theories~\cite{5dAGT1},~\cite{5dAGT2},~\cite{5dAGT3}. In the
Nekrasov--Shatashvili limit~\cite{NS} the resulting equation turns
into the Baxter TQ equation for the quantum XXZ spin chain. Finally,
quantum spectral curve can be thought of as an equation for the
partition function of refined topological strings on toric Calabi-Yau
three-fold with a Lagrangian brane attached to one of the legs of the
toric diagram~\cite{top-strings1},~\cite{top-strings2}. This partition
function can be computed using (refined) topological vertex
technique~\cite{top-vert1},~\cite{top-vert2},~\cite{top-vert3},~\cite{top-vert4}
and in simple cases is related to certain knot
invariants~\cite{Gorsky:2015toa}.

In sec.~\ref{sec:viras-conf-block} we warm up with the well known
example of the ordinary $\beta$-ensemble. In sec.~\ref{sec:q-t-matrix}
we proceed to our main object: the $(q,t)$-matrix model. In
sec.~\ref{sec:ns-limit} we look at the NS limit of the curve and
obtain the Baxter TQ equation for the XXZ spin chain. We also clarify
its relation to the Seiberg--Witten integrable system which in this
case is a given by \emph{another} XXZ spin chain. Some comments on the
obtained result are given in sec.~\ref{sec:concl-disc}.

\section{Dotsenko--Fateev representation and quantum spectral curves}
\label{sec:cft-df-quantum}

\subsection{Virasoro conformal block and $\beta$-ensemble}
\label{sec:viras-conf-block}

Let us start with the usual Virasoro conformal blocks with a single
degenerate field insertion $V_{\mathrm{degen}}$. The differential
equation for this conformal block is obtained from the null-vector
condition $(L_{-1}^2 + a L_{-2}) V_{\mathrm{degen}}(z) = 0$. We will
now translate this statement into the language of $\beta$-ensembles.

Conformal blocks can be represented as correlators in a free field
theory with additional screening operators. This produces the
Dotsenko--Fateev integral representation~\cite{AGTmamo1},~\cite{AGTmamo2},~\cite{AGTmamo3},~\cite{AGTmamo4},~\cite{AGTmamo5},~\cite{MMCB1},~\cite{MMCB2},~\cite{MMCB3}:
\begin{multline}
  \label{eq:1}
  \mathcal{B}_{\mathrm{Vir}} \left(\{ \Lambda_a\}, \{ \Delta_a
    \}\right) = \mathcal{B}_{\mathrm{Heis}}^{-1}\left(\{ \Lambda_a\},
    \{ \Delta_a \}\right) \int_{\{ \mathcal{C}\}} d^N x\,
  \Delta^{2\beta} (x) \prod_{i=1}^N
  \prod_{a=1}^K \left( \Lambda_a - x_i \right)^{v_a}=\\
  = \mathcal{B}_{\mathrm{Heis}}^{-1}\left(\{ \Lambda_a\}, \{ \Delta_a
    \}\right) I_{\mathrm{DF}}\left(\{ \Lambda_a\}, \{ \Delta_a \}, \{
    \mathcal{C} \}\right),
\end{multline}
where $\{ \mathcal{C} \}$ is a collection of contours, which
determines the intermediate dimensions in the conformal block,
$\Delta^{2\beta}(x) = \prod_{i \neq j} (x_i - x_j)^{\beta} $,
$\Lambda_a$ are the positions of the primary fields, and $v_a$ are
Liouville momenta of the primaries. Notice a factor
$\mathcal{B}_{\mathrm{Heis}}^{-1}$, which is the conformal block of
the Heisenberg algebra. It is also known as the ``$U(1)$ part'' in the
AGT literature and is given by a simple explicit expression in
$\Lambda_a$ and $v_a$. From now on we will work exclusively with the
integral itself and not concern ourselves with the Heisenberg part.

To derive the differential equation one needs a primary field with
$v_a = 1$. We denote the position of this field by $z$. Consider the
following integral of a total derivative:
\begin{multline}
  \label{eq:2}
  0 = \int_{\{ \mathcal{C}\}} d^N x\, \sum_{i=1}^N
  \frac{\partial}{\partial x_i} \left[ \frac{1}{z - x_i}
    \Delta^{2\beta} (x) \prod_{k=1}^N (z - x_k) \prod_{i=1}^N
    \prod_{a=1}^K \left( \Lambda_a - x_i \right)^{v_a} \right]=\\
  = \int_{\{ \mathcal{C}\}} d^N x\, \prod_{k=1}^N (z - x_k)
  \sum_{i=1}^N \frac{1}{z - x_i} \frac{\partial}{\partial x_i} \left[
    \Delta^{2\beta} (x) \prod_{i=1}^N \prod_{a=1}^K \left( \Lambda_a -
      x_i \right)^{v_a} \right]=\\
  = \int_{\{ \mathcal{C}\}} d^N x\, \prod_{k=1}^N (z - x_k)
  \sum_{i=1}^N \frac{1}{z - x_i} \left( \sum_{j \neq i} \frac{2
      \beta}{x_i - x_j} - \sum_{a=1}^K \frac{v_a}{\Lambda_a - x_i}
  \right) \left[ \Delta^{2\beta} (x) \prod_{i=1}^N \prod_{a=1}^K
    \left( \Lambda_a - x_i \right)^{v_a} \right] = \\
  = \left( \beta \partial_z^2 - \sum_{a=1}^K \frac{v_a}{z -
      \Lambda_a} \partial_z + \sum_{a=1}^K \frac{1}{z -
      \Lambda_a} \partial_{\Lambda_a} \right)
  I_{\mathrm{DF}}\left(\{z, \Lambda_a\}, \{ \Delta_a \}, \{
    \mathcal{C} \}\right)
\end{multline}
We thus obtain a well-known second order differential equation in $z$
for the conformal block with a single degenerate field. We now move on
to the $q$-deformed case.

\subsection{$q$-Virasoro conformal block and $(q,t)$-matrix model}
\label{sec:q-t-matrix}
In the $q$-deformed case the DF integral becomes Jackson
$q$-integral\footnote{Jackson integral is defined as follows:
  $\int_0^a f(x) d_q x = (1-q) \sum_{n \geq 0} a q^n f(q^n a) =
  \frac{1-q}{1-q^{a \partial_a}} a f(a)$.}:
\begin{equation}
  \label{eq:4}
  I_{\mathrm{DF}} \left( \{\Lambda_a\}, \{v_a\}, \{\mathcal{C}\} \right) = \int_{\{\mathcal{C}\}} d_q^N x\, \Delta^{(q,t)}(x) \prod_{i=1}^N \prod_{a=1}^K
  \prod_{k=0}^{v_a -1} \left( \Lambda_a  - q^k x_i \right),
\end{equation}
where $\{ \mathcal{C} \}$ is a collection of contours,
$\Delta^{(q,t)}(x) = \prod_{k=0}^{\beta-1} \prod_{i \neq j} (x_i - q^k
x_j)$, $t = q^{\beta}$ and $v_{a}, \Lambda_a$ are the
parameters\footnote{We implicitly assume $\beta$ and $v_a$ to be
  integer in these expressions, however everything has a well-defined
  analytic continuation to non-integer values. In particular, the
  spectral curve equation is valid for any values of the parameters.}.

Naturally the quantum spectral curve should now be a difference
equation. The degenerate field insertion in the $q$-deformed case
stays the same: it amounts to adding $\prod_{i=1}^N (z - x_i )$ into
the integral. The construction of the degenerate vertex operator
$V_{1,2}(z)$ associated to the degenerate state $|\alpha_{1,2}\rangle$
was given in~\cite{q-DF1},~\cite{q-DF2},~\cite{q-DF3}. This state
satisfies the $q$-deformed degeneracy condition $(T_{-1}^2 - a_q
T_{-2}) |\alpha_{1,2}\rangle$, where $T_n$ are the generators of the
$q$-deformed Virasoro algebra and $a_q$ is a certain function of $q$
and $t$. The derivatives $\partial_z$ in the quantum spectral curve
equation should be deformed into $q$-derivatives $\left( 1 -
  q^{z \partial_z} \right)$.

As in the previous section, consider the $q$-integral of a total
$q$-derivative:
\begin{multline}
  \label{eq:5}
  0 = \int_{\{\mathcal{C}\}} d_q^N x\, \sum_{i=1}^N \frac{1}{x_i}
  (q^{x_i \partial_{x_i}} - 1 ) \Biggl[\frac{x_i \prod_{a=1}^K
    \left(\Lambda_a - \frac{x_i}{q}\right)}{z - x_i} \prod_{j \neq i}
  \frac{x_i
    - t x_j}{x_i - x_j}\times\\
  \times\Delta^{(q,t)}(x) \prod_{m=1}^N (z - x_m) \prod_{a=1}^K
  \prod_{k=0}^{v_a -1} \left( \Lambda_a -  q^k x_m \right)\Biggr]=\\
  = \int_{\{\mathcal{C}\}} d_q^N x\, \prod_{k=1}^N (z - x_k)
  \sum_{i=1}^N \frac{1}{z - x_i} (q\, q^{x_i \partial_{x_i}} - 1 )
  \Biggl[\prod_{a=1}^K \left(\Lambda_a - \frac{x_i}{q}\right) \prod_{j
    \neq i} \frac{x_i
    - t x_j}{x_i - x_j}\times\\
  \times \Delta^{(q,t)}(x)\prod_{m=1}^N \prod_{a=1}^K \prod_{k=0}^{v_a -1} \left(
    \Lambda_a - q^k x_m \right)\Biggr]
\end{multline}
Notice a nontrivial factor of $\prod_{j \neq i} \frac{x_i - t x_j}{x_i
  - x_j}$, which is inherent to the $q$-deformed case and does not
appear in the limit $q \to 1$. The same factor appears in Macdonald
operator, of which Macdonald polynomials are eigenfunctions. However,
the connection of this operator and the quantum spectral curve is not
yet clear. The shift operator $q^{x_i \partial_{x_i}}$ acts on
$\Delta^{(q,t)}(x)$ as follows:
\begin{equation}
  \label{eq:6}
  q^{x_i \partial_{x_i}} \Delta^{(q,t)}(x) = t^{N-1} \prod_{j \neq i} \frac{(q x_i - x_j) (t x_i -
    x_j)}{(x_i - x_j) (q x_i - t x_j )} \Delta^{(q,t)}(x).
\end{equation}
Some factors in Eq.~\eqref{eq:6} cancel with the additional factor
$\prod_{j \neq i} \frac{x_i - t x_j}{x_i - x_j}$. One obtains the
following identity:
\begin{multline}
  \label{eq:7}
  0 = \int_{\{\mathcal{C}\}} d_q^N x\, \Biggl[ \sum_{i=1}^N \frac{q
    t^{N-1}}{z - x_i} \prod_{j \neq i} \frac{t x_i - x_j}{x_i - x_j}
  \prod_{a=1}^K \left(\Lambda_a  - q^{v_a} x_i\right)-\\
  - \sum_{i=1}^N \frac{1}{z - x_i} \prod_{j \neq i} \frac{x_i - t
    x_j}{x_i - x_j} \prod_{a=1}^K \left(\Lambda_a -
    \frac{x_i}{q}\right) \Biggr] \times\\
  \times \Delta^{(q,t)}(x) \prod_{k=1}^N (z - x_k) \prod_{a=1}^K
  \prod_{k=0}^{v_a -1} \left( \Lambda_a - q^k x_i \right).
\end{multline}
Let us write both terms in the square brackets as contour
integrals\footnote{We include the factor $\frac{1}{2\pi i}$ in the
  definition of the contour integral.}:
\begin{multline}
  \label{eq:8}
  \sum_{i=1}^N \frac{q t^{N-1}}{z - x_i} \prod_{j \neq i} \frac{t x_i
    - x_j}{x_i - x_j} \prod_{a=1}^K \left(\Lambda_a -
    q^{v_a}  x_i\right) =\\
  =\frac{q t^{N-1}}{t-1} \oint_{\mathcal{C}_x} \frac{dw}{w} \frac{1}{z
    - w} \prod_{j=1}^N \frac{t w - x_j}{w - x_j} \prod_{a=1}^K \left(
    \Lambda_a - q^{v_a} w \right),
\end{multline}
\begin{multline}
\sum_{i=1}^N \frac{1}{z - x_i} \prod_{j \neq i} \frac{x_i - t
    x_j}{x_i - x_j} \prod_{a=1}^K \left(\Lambda_a -
    \frac{x_i}{q}\right) =\\
  =\frac{1}{1-t} \oint_{\mathcal{C}_x} \frac{dw}{w} \frac{1}{z - w}
  \prod_{j=1}^N \frac{w - t x_j}{w - x_j} \prod_{a=1}^K \left(
    \Lambda_a - \frac{w}{q} \right), \label{eq:9}
\end{multline}
where the contour $\mathcal{C}_x$ encircles all the points $x_i$.

We deform the contour $\mathcal{C}_x$ to $\mathcal{C}_{\infty}$ --- a
large contour around zero --- so that it gets extra contributions from
residues at $z$ and $0$. We will treat the integral over
$\mathcal{C}_{\infty}$ in a moment. Let us first find the residues at
points $z$ and $0$:
\begin{multline}
  \label{eq:10}
  \frac{q t^{N-1}}{t-1} \oint_{\mathcal{C}_x} \frac{dw}{w} \frac{1}{z
    - w} \prod_{j=1}^N \frac{t w - x_j}{w - x_j} \prod_{a=1}^K \left(
    \Lambda_a - q^{v_a} w \right) =\\
  = - \frac{q t^{N-1}}{z(t-1)} \left[ \prod_{a=1}^K \Lambda_a -
    \prod_{j=1}^N \frac{tz - x_j}{z - x_j} \prod_{a=1}^K \left(
      \Lambda_a - q^{v_a} z \right) \right],
\end{multline}
\begin{multline}
  \frac{1}{1-t} \oint_{\mathcal{C}_x} \frac{dw}{w} \frac{1}{z - w}
  \prod_{j=1}^N \frac{w - t x_j}{w - x_j} \prod_{a=1}^K \left(
    \Lambda_a - \frac{w}{q} \right) = \\
  = \frac{1}{z(t-1)} \left[ t^N \prod_{a=1}^K \Lambda_a -
    \prod_{j=1}^N \frac{z - tx_j}{z - x_j} \prod_a \left( \Lambda_a -
      \frac{z}{q} \right) \right]. \label{eq:11}
\end{multline}
One can notice that the factors containing $x_j$ in
Eqs.~\eqref{eq:10},~\eqref{eq:11} can be expressed as shift operators
acting on the degenerate field:
\begin{gather}
  \label{eq:12}
  \prod_{j=1}^N \frac{tz - x_j}{z - x_j} \prod_{j=1}^N (z - x_j) =
  t^{z \partial_z} \prod_{j=1}^N (z - x_j),\\
  \prod_{j=1}^N \frac{z - t x_j}{z - x_j} \prod_{j=1}^N (z - x_j) =
  t^N t^{-z \partial_z} \prod_{j=1}^N (z - x_j). \label{eq:13}
\end{gather}

We now calculate the troublesome residues at infinity:
\begin{multline}
  \label{eq:14}
  \frac{q t^{N-1}}{t-1} \oint_{\mathcal{C}_{\infty}} \frac{dw}{w}
  \frac{1}{z - w} \prod_{j=1}^N \frac{t w - x_j}{w - x_j}
  \prod_{a=1}^K \left(
    \Lambda_a - q^{v_a}  w \right) = \\
  =\frac{q t^{N-1}}{(t-1)}\frac{1}{(K-1)!} \frac{\partial^{K-1}}{\partial
    y^{K-1}} \left.\left( \frac{1}{yz-1} \prod_{j=1}^N \frac{t - y
        x_j}{1 - y x_j} \prod_{a=1}^K \left( \Lambda_a y - q^{v_a} \right) \right) \right|_{y=0} ,
\end{multline}
\begin{multline}
  \label{eq:15}
  \frac{1}{1-t} \oint_{\mathcal{C}_{\infty}} \frac{dw}{w} \frac{1}{z -
    w} \prod_{j=1}^N \frac{w - t x_j}{w - x_j} \prod_{a=1}^K \left(
    \Lambda_a - \frac{w}{q} \right) = \\
  = \frac{1}{q^K (1-t)}\frac{1}{(K-1)!}
  \frac{\partial^{K-1}}{\partial y^{K-1}} \left.\left( \frac{1}{yz-1}
      \prod_{j=1}^N \frac{1 - t y x_j}{1 - y x_j} \prod_{a=1}^K \left(
        \Lambda_a q y - 1 \right) \right) \right|_{y=0} ,
\end{multline}
For lower $K$ the total residue at infinity, which we denote by
$R^{(K)}_{\infty}$, can be found explicitly:
\begin{align}
  R^{(0)}_{\infty} &= 0, \notag\\
  R^{(1)}_{\infty} &= \frac{1}{q(t-1)}\left( t^{2N-1}
    q^{v_1+2} - 1\right), \notag\\
  R^{(2)}_{\infty} &= \frac{1}{q^2(t-1)} \Biggl[ z \left( t^{2N-1}
    q^{3+v_1 +v_2} - 1 \right) + (t-1) \left( q^{3+ v_1+ v_2}
    t^{2N-2} + 1 \right) \sum_{j=1}^N x_j + \notag\\
  &\phantom{=} + \sum_{a=1}^2\Lambda_a q^{-v_a} \left(q^{1+ v_a} -
    q^{3+ v_1 + v_2} t^{2N-1}\right) \Biggr] \notag
\end{align}
In general $R^{(K)}_{\infty}(z)$ is a polynomial in $z$ of degree
$K-1$, of the form
\begin{equation}
  \label{eq:18}
  R^{(K)}_{\infty}(z) =
  \frac{t^{2N-1}q^{1+\sum_{a=1}^K v_a} - q^{-K}}{t-1} (-z)^{K-1} + \ldots
\end{equation}
The coefficient in front of $z^m$ contains symmetric polynomials in
$x_j$ up to degree $K-m-1$. In the NS limit these symmetric
polynomials decouple from the matrix model average and become
Hamiltonians of the XXZ spin chain. Before the limit, they can be
obtained by acting on $I_{\mathrm{DF}}(z)$ with certain operators in
$\Lambda_a$. We denote the operator polynomial, which produces
$R^{(K)}_{\infty}(z)$ from $I_{\mathrm{DF}}(z)$ as
$\hat{R}^{(K)}_{\infty}(z)$. However, we will not write out these
operators explicitly.

Finally, we obtain the quantum spectral curve for the $(q,t)$-matrix
model:
\begin{multline}
  \label{eq:16}
  \Biggl[ - \frac{t^N}{z(t-1)}\left( \left(\frac{q}{t} - 1 \right)
    \prod_{a=1}^K \Lambda_a - \frac{q}{t} \prod_{a=1}^K \left(
      \Lambda_a - q^{v_a} z \right) t^{z \partial_z} + \prod_{a=1}^K
    \left( \Lambda_a - \frac{z}{q}
    \right) t^{-z \partial_z} \right) +\\
  +  \hat{R}^{(K)}_{\infty}(z) \Biggr]
  I_{\mathrm{DF}}(z) = 0.
\end{multline}

One usually sets the position of one of the fields in the conformal
block to zero, which simply amounts to taking the limit $\Lambda_1 \to
0$. Then the equation simplifies:
\begin{equation}
  \label{eq:17}
\boxed{  \left[ \frac{t^N}{(t-1)}\left(\frac{q^{v_1 + 1}}{t} \prod_{a=2}^K \left( \Lambda_a - q^{v_a} z \right)
      t^{z \partial_z} - \frac{1}{q} \prod_{a=2}^K \left( \Lambda_a - \frac{z}{q}
      \right) t^{-z \partial_z} \right) + \hat{R}^{(K)}_{\infty}(z) \right] I_{\mathrm{DF}}(z) = 0}
\end{equation}

\section{Nekrasov--Shatashvili limit}
\label{sec:ns-limit}

In the NS limit the quantum spectral curve can be interpreted as a
Baxter equation for a certain integrable system~\cite{BS1},~\cite{BS2},~\cite{BS3},~\cite{BS4},~\cite{BS5},~\cite{BS6} (for a more
mathematical description see~\cite{math1}~\cite{math2}). In the
ordinary Virasoro case this system is just the Hitchin integrable
system associated to the Riemann surface (with marked points) on which
the conformal block lives. Generalizing this result, we show that the
$q$-deformed conformal block on a sphere satisfies the Baxter TQ
equation for the XXZ spin chain.

\subsection{Virasoro block and Hitchin system}
\label{sec:viras-hitch-syst}

Again we start with the well-known example of Virasoro conformal
block. In the NS limit $\epsilon_2 \to 0$, $\epsilon_1 =
\mathrm{const}$\footnote{The parameters $\epsilon_1$ and $\epsilon_2$
  enter Nekrasov function on equal footing, so there is an equivalent
  form of the NS limit $\epsilon_1 \to 0$, $\epsilon_2 =
  \mathrm{const}$, which we employ in the next section.} so that
$\beta \to \infty$, and the dimensions of the fields in the conformal
block also have to be rescaled with $\beta$, so that $v_a = \beta
\widetilde{v}_a$, $\widetilde{v}_a = \mathrm{const}$. The main
simplification in the quantum spectral curve equation~\eqref{eq:2}
occurs because the dimension of the degenerate field is fixed and does
not scale with $\beta$. Therefore, the correlators involving
$V_{\mathrm{degen}}(z)$, in particular $I_{\mathrm{DF}}(z)$, behave as
$e^{-\beta \mathcal{F} + \mathcal{S}(z) + \ldots}$, where the leading
factor $\mathcal{F}$ is independent of $z$. To get rid of the leading
factor one can divide by the DF integral $I_{\mathrm{DF},0} \simeq
e^{-\beta \mathcal{F}}$ \emph{without} the degenerate field insertion.

Finally, we obtain an equation for the ``wave function'' $\Psi (z) =
\frac{I_{\mathrm{DF}}(\{ z, \Lambda_a \})}{I_{\mathrm{DF},0}(\{
  \Lambda_a \})}$:
\begin{equation}
  \label{eq:3}
  \left( \partial_z^2 - \sum_{a=1}^K \frac{\widetilde{v}_a}{z -
      \Lambda_a} \partial_z + \sum_{a=1}^K \frac{h_a}{z -
      \Lambda_a} \right) \Psi(z) = 0
\end{equation}
where $h_a = \partial_{\Lambda_a} \mathcal{F}(\{\Lambda_a\})$. This is
nothing but the Baxter equation for the $\mathfrak{gl}_2$ Gaudin model
with $K$ marked points which is the relevant Hitchin system in this
case. The Baxter equation can be written in the form
\begin{equation}
  \label{eq:19}
  :\det_{2\times 2} (\partial_z - L(z) ): \Psi(z) =0,
\end{equation}
where $L(z) = \sum_{a=1}^K \frac{A_a}{z-\Lambda_a}$ is the Lax matrix
and an appropriate normal ordering is imposed on the operators in the
determinant. The parameters of the $\beta$-ensemble are interpreted as
the Casimir operators and Hamiltonians of the Gaudin model as follows:
\begin{align}
  \label{eq:20}
  \widetilde{v}_a &= \tr A_a,\\
  h_a &= \sum_{b \neq a} \frac{\tr A_a A_b}{\Lambda_a -
    \Lambda_b}.\label{eq:21}
\end{align}

\subsection{$q$-deformed block and XXZ spin chain}
\label{sec:q-deformation-xxz}
In the $q$-deformed case one has $q = e^{-R_5 \epsilon_1}$, $t = e^{R_5
  \epsilon_2}$ and in the NS limit $\epsilon_1 \to 0$, $\epsilon_2 =
\mathrm{const}$. This choice is related to the convention of the
previous section by the symmetry $\epsilon_1 \leftrightarrow
\epsilon_2$. One also has to rescale the $v_a$ parameters so that
$q^{v_a} = t^{\widetilde{v}_a} = \mathrm{const}$.

As in the ordinary Virasoro case, in the NS limit the operator
$\hat{R}^{(K)}_{\infty}(z)$ act on the DF integral without insertion
$I_{\mathrm{DF},0}$ and not on the wavefunction $\Psi(z) =
\frac{I_{\mathrm{DF}}(\{ z, \Lambda_a \})}{I_{\mathrm{DF},0}(\{
  \Lambda_a \})}$. Thus the operator $\hat{R}^{(K)}_{\infty}(z)$ is
replaced by a polynomial $\frac{t^N}{t-1} P_K(z) = \left( I_{\mathrm{DF},0}^{-1}(\{
  \Lambda_a \}) \hat{R}^{(K)}_{\infty}(z) I_{\mathrm{DF},0}(\{
  \Lambda_a \})\right) $ and the quantum spectral curve~\eqref{eq:17}
takes the form
\begin{equation}
  \label{eq:22}
  \left[ t^{\widetilde{v}_1 - 1} \prod_{a=2}^K \left( \Lambda_a - t^{\widetilde{v}_a} z \right)
    t^{z \partial_z} - \prod_{a=2}^K \left( \Lambda_a - z
    \right) t^{-z \partial_z} + P_{K-1}(z) \right] \Psi(z) = 0
\end{equation}

This equation has exactly the same form as the Baxter equation for the
XXZ spin chain with $K-1$ spins from $U_q(\mathfrak{gl}_2)$algebra:
\begin{equation}
  \label{eq:24}
  :\det_{2 \times 2} (q^{z\partial_z} - T(z) ): Q(z) = 0
\end{equation}
where $T(z) = V \prod_{a=2}^K L_a(z)$ is the transfer matrix. More
explicitly
\begin{equation}
  \label{eq:23}
  \left[ \mathfrak{q} K_{+}(z) q^{z \partial_z} - K_{-}(z) q^{-z \partial_z} + P_{K-1}(z) \right] Q(z) = 0,
\end{equation}
where $K_{\pm}(z) = \prod_{a=2}^K (z - m_a^{\pm})$ and $P_{K-1}(z) =
\sum_{a=0}^{K-1} h_a z^{K-a-1}$ are polynomials of degree $K-1$
containing Casimir operators of the spins and the Hamiltonians of the
chain respectively and $\mathfrak{q}$ is the parameter of the twist
matrix. The matrix model parameters are identified with the spin chain
parameters as follows:
\begin{align}
  m_a^{+} &= \Lambda_a t^{-\widetilde{v}_a}, \label{eq:25}\\
  m_a^{-} &= \Lambda_a , \label{eq:26}\\
  \mathfrak{q} &= t^{\sum_{a=1}^K \widetilde{v}_a -1} \label{eq:27}
\end{align}
and one should write $q$ instead of $t$ in Eq.~\eqref{eq:22}.

Let us clarify the relation of the integrable system we have just
obtained and the Seiberg--Witten integrable
system~\cite{IntSyst1},~\cite{IntSyst2}. Due to the AGT correspondence
the $q$-deformed conformal block we consider is equal to partition
function of the five-dimensional $SU(2)^{K-2}$ gauge theory. In turn
this gauge theory is associated with the XXZ spin chain with two spins
from symmetric representations of $U_q(\mathfrak{gl}_{K-1})$
algebra. This is notably \emph{different} from the spin chain
described by~\eqref{eq:23}, which has $K-1$ spins of
$U_q(\mathfrak{gl}_2)$.

The resolution of this seeming discrepancy requires the notion of
\emph{spectral duality}. There are in fact two equivalent chains with
the Baxter equations related by this duality (it can be thought of as
just the Fourier transform in the spectral
parameter)~\cite{spectral1},~\cite{spectral2},~\cite{spectral3},~\cite{spectral4}. One
is the Seiberg--Witten integrable system, while the other is the
system arising from the AGT dual conformal block. The two systems are
described by two different $(q,t)$-matrix models, which are related by
a very nontrivial integral transform. To make the spectral duality
explicit one needs to decompose both integrals in terms of refined
topological vertices and use the slicing invariance property of
refined topological strings.

\section{Conclusions and discussion}
\label{sec:concl-disc}

We have derived the quantum spectral curve for the $(q,t)$-matrix
model and shown that in the NS limit it reduces to the Baxter equation
for the XXZ spin chain. Spectral duality connects this spin chain and
the corresponding Seiberg--Witten integrable system. This should be
compared with the quantum spectral curve for the $\beta$-ensemble,
which in the NS limit is related to Hitchin system.

There are several directions in which our result can be
generalized. Firstly, one can consider the toric $q$-deformed
conformal blocks and the corresponding Dotsenko--Fateev
integrals. They should satisfy a difference equation, of the elliptic
Ruijsenaars form. It would also be interesting to study elliptic
deformations of the Virasoro \emph{algebra}, which should lead to XYZ
spin chains and \emph{double elliptic} integrable
models~\cite{Dell1},~\cite{Dell2}. Probably Baxter equations for these
cases might clarify the relationship between the spectral
duality~\cite{spectral1},~\cite{spectral2},~\cite{spectral3},~\cite{spectral4}
and p-q duality in integrable
models~\cite{p-q1},~\cite{p-q2},~\cite{p-q3},~\cite{p-q4}. Recently
the connection between $5d$ gauge theories described by the matrix
models we have considered here and polynomial knot invariants has been
obtained~\cite{Gorsky:2015toa}. This direction is also worth
investigating in the future.

\textbf{Note added in the published version.} While this paper was
refereed a number of other works~\cite{qq1},~\cite{qq2},~\cite{qq3} dealing with
$(q,t)$-deformation of the classical spectral curve have
appeared. These papers consider matrix models which are similar to or
generalizations of the $(q,t)$-matrix model we have considered. They
derive the recurrence relations for the correlators (also known as the
$qq$-characters, loop equations or nonperturbative Ward identities) in
these matrix models, which are equivalent to the $(q,t)$-spectral
curve~(\ref{eq:17}) for any values of $q$ and $t$. The only essential
difference between the two approaches is that the authors of~\cite{qq1},~\cite{qq2},~\cite{qq3}
consider an insertion of the energy-momentum tensor $T(z)$ in the
matrix model average, while in the present work an insertion of the
product $T(z) V_{1,2}(z)$ of the energy-momentum tensor with a
degenerate vertex operator $V_{1,2}(z)$ is considered.

\textbf{Acknowledgements.} The author is grateful to D.~Galakhov and
T.~Kimura for attracting his attention to the problem. The author
appreciates the hospitality and stimulating atmosphere of the Lorentz
Center, Leiden, Netherlands, where part of this work was done. The
author is supported by the RFBR grants 17-01-00585, 16-51-45029-INDa,
15-51-52031-NSC-a, 16-51-53034-GFEN, 17-51-50051-YaF, by INFN, by the
ERC Starting Grant 637844-HBQFTNCER and by D. Zimin's ``Dynasty''
foundation stipend.

\end{document}